\documentclass[journal=jacsat,manuscript=article]{achemso}

\usepackage[version=3]{mhchem} 

\usepackage{xcolor}

\author{Frederico B. Sousa\footnotemark[1]\footnotetext{$^\ast$ F.B.S. and R.B.N. contributed equally to this paper}}
\affiliation{Departamento de F\'isica, Universidade Federal de Minas Gerais, Belo Horizonte, Minas Gerais 30123-970, Brazil}

\author{Rafael Battistella Nadas\footnotemark[1]}
\affiliation{Departamento de F\'isica, Universidade Federal de Minas Gerais, Belo Horizonte, Minas Gerais 30123-970, Brazil}

\author{Rafael Martins}
\affiliation{Departamento de F\'isica, Universidade Federal de Ouro Preto, Ouro Preto, Minas Gerais 35400-000, Brazil}

\author{Ana P. M. Barboza}
\affiliation{Departamento de F\'isica, Universidade Federal de Ouro Preto, Ouro Preto, Minas Gerais 35400-000, Brazil}

\author{Jaqueline S. Soares}
\affiliation{Departamento de F\'isica, Universidade Federal de Ouro Preto, Ouro Preto, Minas Gerais 35400-000, Brazil}

\author{Bernardo R. A. Neves}
\affiliation{Departamento de F\'isica, Universidade Federal de Minas Gerais, Belo Horizonte, Minas Gerais 30123-970, Brazil}

\author{Ive Silvestre}
\affiliation{Departamento de F\'isica, Universidade Federal de Ouro Preto, Ouro Preto, Minas Gerais 35400-000, Brazil}

\author{Ado Jorio}
\affiliation{Departamento de F\'isica, Universidade Federal de Minas Gerais, Belo Horizonte, Minas Gerais 30123-970, Brazil}

\author{Leandro M. Malard}
\affiliation{Departamento de F\'isica, Universidade Federal de Minas Gerais, Belo Horizonte, Minas Gerais 30123-970, Brazil}
\email{lmalard@fisica.ufmg.br}

\title[An \textsf{achemso} demo]
  {Nano-optical investigation of grain boundaries, strain and edges in CVD grown MoS$_{2}$ monolayers}

\abbreviations{IR,NMR,UV}

\begin{document}



\begin{abstract}

The role of defects in two-dimensional semiconductors and how they affect the intrinsic properties of these materials have been a wide researched topic over the past decades. Optical characterization such as photoluminescence and Raman spectroscopies are important tools to probe their physical properties and the impact of defects. However, conventional optical techniques present a spatial resolution limitation lying in a $\mu$m-scale, which can be overcomed by the use of near-field optical measurements. Here, we use tip-enhanced photoluminescence and Raman spectroscopies to unveil nanoscale optical heterogeneities at grain boundaries, local strain fields and edges in grown MoS$_{2}$ monolayers. A noticeable enhancement of the exciton peak intensity corresponding to a trion emission quenching is observed at narrow regions down to 47 nm of width at grain boundaries related to doping effects. Besides, localized strain fields inside the sample lead to non-uniformities in the intensity and energy position of photoluminescence peaks. Finally, distinct samples present different nano-optical responses at their edges due to strain and passivation defects. The passivated defective edges show a photoluminescence intensity enhancement and energy blueshift as well as a frequency blueshift of the 2LA Raman mode. On the other hand, the strained edges display a photoluminescence energy redshift and frequency redshifts for E$_{2g}$ and 2LA Raman modes. Our work shows that different defect features can be only probed by using optical spectroscopies with a nanometric resolution, thus revealing hindered local impact of different nanoscale defects in two-dimensional materials.

\end{abstract}


Two-dimensional transition metal dichalcogenides (TMD) are semiconducting van der Waals materials with remarkable optical and electronic properties such as a transition from an indirect to a direct band gap when thinned to a monolayer \cite{Mak2010,Splendiani2010}, strong many-body effects \cite{Mak2013,Chernikov2014,Ugeda2014,He2014,Sousa2023} and circular polarized light with valley selectively \cite{Xiao2012,Mak2012,Zeng2012,Cao2012}. These properties lead TMDs as potential materials for the next generation of opto-electronic devices \cite{Radisavljevic2011,Lopez-Sanchez2013,Pospischil2014}. However, the grown TMD samples with the large-area required for industrial-scale applications commonly present defects as vacancies, strain fields, electrical and chemical doping, grain boundaries and surface adsorbents \cite{Lin2016}. These defects are related with optical and electronic properties modifications, which can deeply affect the TMD performance in devices \cite{Zhang2019}. On the other hand, there is also a great effort in the defect engineering of these 2D TMDs in order to achieve novel desirable properties \cite{Lin2016,Liang2021}. Therefore, the characterization of these defects in TMDs is essential to both help in the development of their growth methods as well as to study the intentionally tuning of their properties.

These defects can be well ascertained with a nanoscale spatial resolution by scanning microscopy measurements such as scanning probe microscopy (SPM), scanning tunneling microscopy (STM) and scanning transmission electron microscopy (STEM) \cite{Addou2015,Prado2015,Fei2016,Wang2018-2}. However, to investigate the role of these defects in the inhomogeneity of the optical properties throughout the grown samples there is a spatial resolution limitation of hundreds of nanometers due to the diffraction limit of light. Therefore, confocal optical techniques as $\mu$-photoluminescence (PL) and $\mu$-Raman spectroscopy are only able to provide averaged electronic and vibrational responses over a $\mu$m-scale area. Hence, the precise determination of localized optical modifications by defects cannot be resolved by these techniques. Recently, tip-enhanced PL and Raman spectroscopies (TEPL and TERS, respectively) have also been used to study defective optical features in TMDs \cite{Malard2021,kim2021near,Shao2022,Kwon2023}. As the tip amplifies the light signal due to its plasmonic field, the spatial resolution of these tip-enhanced techniques is determined by the tip size. Thus, investigations of excitonic effects \cite{Su2016,Okuno2018}, grain boundaries \cite{Lee2015,Smithe2018,Su2021}, edge defects \cite{Rodriguez2019,Huang2019,Wang2023}, strain \cite{Rahaman2017}, wrinkles \cite{Kato2019,Shao2022-2}, lateral interfaces \cite{Sahoo2019,Shao2021,Fali2021,Garg2022}, hybrid heterostructures \cite{Gadelha2023} and defect-bound localized states \cite{Lee2017,Wang2023} have been carried out by TEPL and TERS measurements. However, due to the implementation challenges of these optical techniques, it is possible to observe that several of these works have not reached the optimal spatial resolution of few dozens of nanometers. Moreover, as defects in grown TMDs can show distinct optical modifications depending on further factors as growth parameters and substrate \cite{Buscema2014,Senkic2023}, a wide nanoscale characterization with combined electronic and vibrational information of different defects in TMD grown monolayers is still a demanding task.

In this work we studied several defects in chemical vapor deposition (CVD) grown MoS$_{2}$ monolayers by TEPL and TERS measurements with spatial resolutions down to $\sim$$24$ nm. We investigated the PL emission along distinct MoS$_{2}$ monolayers grain boundaries, revealing a remarkable localized intensity enhancement of the exciton peak due to a suppression of trion formation along these defects. Furthermore, non-uniform nanoscale strain fields throughout the sample were also probed by TEPL and TERS measurements. Finally, MoS$_{2}$ monolayers from distinct set of samples presented different defective edge features. In one case, the MoS$_{2}$ monolayer displayed an enhanced and blueshifted PL signal at the edges. This PL shift is associated with a frequency shift of $4$ cm$^{-1}$ in the second-order 2LA Raman mode and not at the first order Raman modes, implying in a possible defective electronic state. On the other hand, MoS$_{2}$ monolayers from the other sample set revealed a distinct response in a $50$ nm width region at their edges. A PL redshift is observed at the edges, that corresponds to remarkable frequency redshifts for E$_{2g}$ and 2LA Raman modes. Atomic force microscopy (AFM) measurements were also performed and showed a suspension of $2$ nm at the edges, which suggests the presence of a strain field at the MoS$_{2}$ monolayers edges. Our results highlights the importance of probing both electronic and vibrational responses with nanometer resolution in the 2D materials.

\section{Results and discussion}

The MoS$_{2}$ monolayer samples A1 and A2 investigated here were grown by a CVD method in two different batches that led to two different optical characteristics as we will show along the manuscript. Further growth details are described in the Methods Section. We first studied a multi-sided polygonal shaped MoS$_{2}$ monolayer of A1 sample (see Supporting Information Figure S1a for its optical image). It is well-known that these CVD grown multi-sided TMD monolayers present grains with distinct crystallographic orientations \cite{VanderZande2013,Zhou2013}. The boundary between the two grains is thus a defective region in which the mismatched atoms present a distinct localized geometrical structure that depends on the relative orientation of the grains. These grain boundaries have been studied for different TMD monolayers in the past years \cite{VanderZande2013,Zhou2013,Azizi2014,Ly2014,Kim2016,Karvonen2017,Rosa2022}. While SPM measurements are capable of determining the nanoscale features of the modified atomic structure along the grain boundaries \cite{VanderZande2013,Zhou2013,Azizi2014}, conventional optical measurements only give the averaged response of $\mu$m-areas containing these defects \cite{VanderZande2013,Ly2014,Kim2016,Karvonen2017,Rosa2022}. Therefore, although intensity enhancements and peak shifts had been reported for TMD monolayers grain boundaries by $\mu$-PL and $\mu$-Raman measurements, they were not able to ascertain the area in which the optical properties are affected due to their limited spatial resolution. More recently, TEPL and TERS measurements have also been carried out in monolayer TMDs in order to provide a wider comprehension of the nanoscale character of the grain boundaries optical features \cite{Lee2015,Smithe2018,Su2021}. However, these few existent reports showed quenched Raman and PL intensities along the grain boundaries, contrary to previously reported features and in accordance with a likely sample damage at these defects. Hence, a careful nano-optical investigation of TMD grain boundaries is required for a deeper understanding of the roles of defects in monolayer TMDs.

\begin{figure}[!htb]
 \centering
 \includegraphics[scale=0.35]{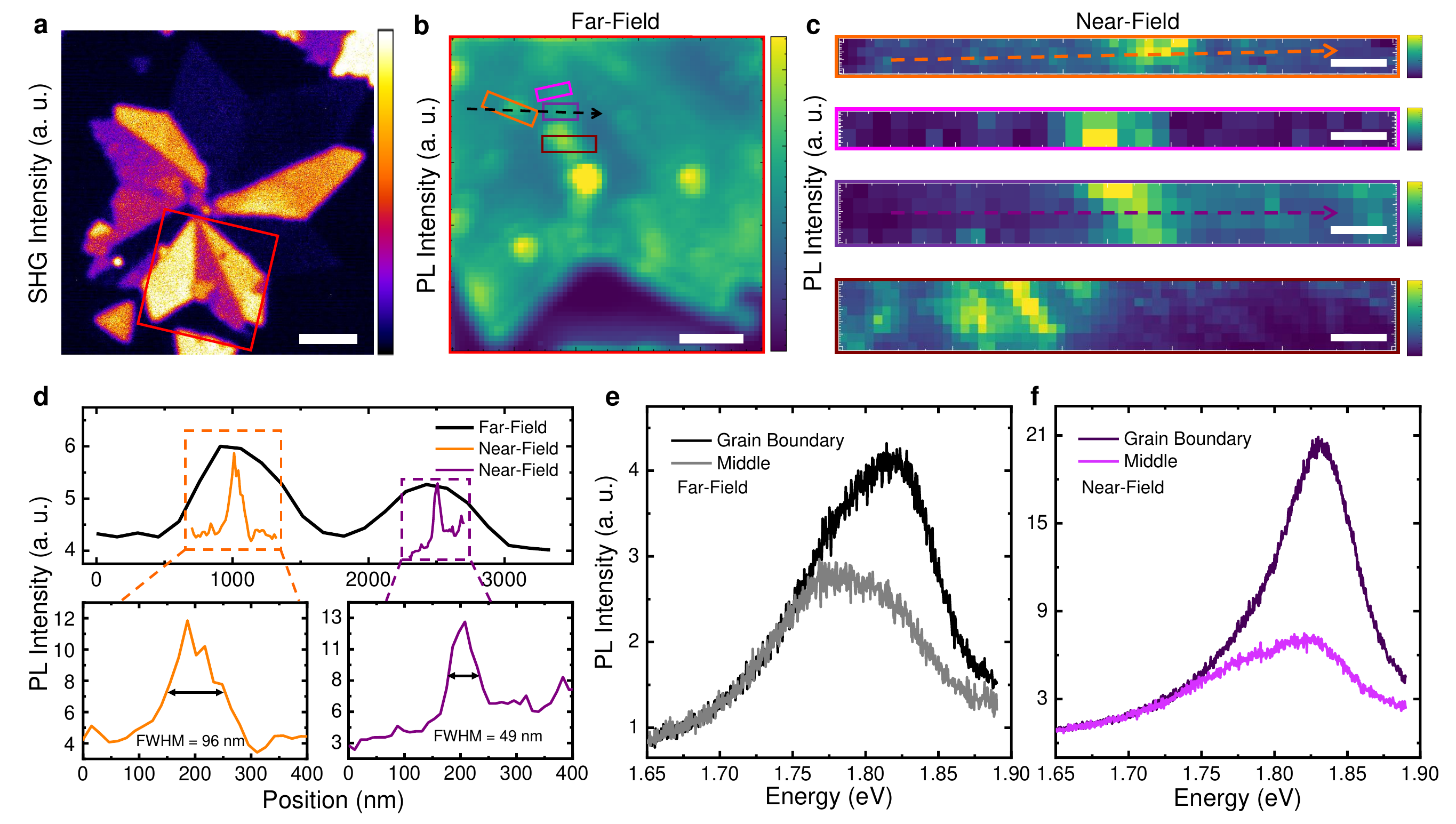} 
 \caption{{\small {\bf a} Polarized SHG imaging of a multi-sided polygonal MoS$_{2}$ monolayer. The regions with distinct SHG intensities represent grains with different crystallographic orientations, showing all grain boundaries present in the sample. Scale bar in ({\bf a}): 5 $\mu$m. {\bf b} PL intensity map of a FF hyperspectral measurement taken in the red squared area highlighted in ({\bf a}). The grain boundary regions show a PL intensity enhancement. Scale bar in ({\bf b}): 2 $\mu$m. {\bf c} TEPL intensity maps of 4 NF hyperspectral measurements taken in the colored rectangular areas highlighted in ({\bf b}) also showing a PL intensity enhancement at the grain boundaries. The colors of the rectangular areas in ({\bf b}) are respective to the same underlined colors around the TEPL maps in ({\bf c}). The FF hyperspectral measurements were performed with steps of 150 nm, while NF measurements were carried with steps of 16 nm. Scale bars in ({\bf c}): 50, 100, 40 and 100 nm (from the top to the bottom). {\bf d} PL intensity profiles along grain boundaries for FF and NF measurements. The top graph shows the FF PL intensity profile in black (taken along the black dashed arrow of {\bf b}) and the NF PL intensity profiles in orange and purple (taken along the orange and purple dashed arrows of {\bf c}). The spatial width of the PL enhancement at the grain boundaries are shown in the bottom intensity profile graphs. {\bf e,f} FF and NF PL spectra, respectively, of grain boundary and grain middle regions. The spectra are respective to the grain boundary region shown in the purple squared region of ({\bf b}) and ({\bf c}). NF measurements reveal the real magnitude of the PL enhancement at the grain boundaries, also showing the exciton emission dominance over the trion emission suppression.}}
 \label{fig1}
 \end{figure}

Figure \ref{fig1}a shows a polarized second-harmonic generation (SHG) image of the multi-sided polygonal shaped MoS$_{2}$ monolayer (see Methods Section for experimental details). Since the SHG in monolayer TMDs is sensitive to the material crystallographic orientation \cite{Malard2013}, polarized SHG measurements are capable of determining the grain orientations and thus reveals the presence of grain boundaries \cite{Yin2014}. Hence, the different SHG intensities in Figure \ref{fig1}a shows the distinct grains and their boundaries throughout the MoS$_{2}$ monolayer. Supporting Information Figure S1c displays the entire polarization dependence of the observed grains and their relative crystallographic orientations. Figure \ref{fig1}b shows the intensity map of a $\mu$-PL hyperspectral measurement performed in the red squared region displayed in Figure \ref{fig1}a. In agreement with previously reports \cite{VanderZande2013,Kim2016,Karvonen2017,Rosa2022}, PL intensity enhancements are observed at grain boundaries. Therefore, in order to ascertain the localization and the real magnitude of the optical modifications in these grain boundaries, we performed TEPL measurements along distinct grain boundaries regions highlighted in colored rectangles in Figure \ref{fig1}b. To show the spatial resolution and signal enhancements of the technique, all TEPL and TERS measurements in this work were carried out with the tip up and down, respective to their far-field (FF) and near-field (NF) responses (further experimental details in Methods Section). Figure \ref{fig1}c displays PL intensity maps of 4 TEPL hyperspectral measurements along MoS$_{2}$ monolayer grain boundaries regions. The underlined color around each TEPL hyperspectral map is respective to the rectangular areas of the same color in Figure \ref{fig1}b. A localized PL intensity enhancement is noted for all probed grain boundaries. Moreover, an energy blueshift is also observed at the grain boundaries, as shown in Supporting Information Figure S2. To quantify the spatial width of the optical modifications in these defects probed by the NF measurements and compare them with the spatial resolution of the FF measurements, Figure \ref{fig1}d shows PL intensity profiles taken along the black dashed arrow in Figure \ref{fig1}b and orange and purple dashed arrows in Figure \ref{fig1}c. The top intensity profile graph of Figure \ref{fig1}d reveals the remarkable increase in spatial resolution given by the tip. While the grain boundaries in the FF measurement presents spatial widths of $\sim 650$ nm, the bottom intensity profile graphs unveil that the region affected by these defects can be as narrow as 49 nm. Distinct grain boundaries present different structural modifications, hence it is natural that they also display different affected regions. For instance, the grain boundary presented in the brown underlined TEPL map of Figure \ref{fig1}c shows two separated enhanced lines, that correspond to a broader brighter region in the FF PL map of Figure \ref{fig1}b. Figures \ref{fig1}e,f compare the FF and NF PL spectra, respectively, of grain boundary and grain middle regions. The FF spectra present a PL energy blueshift and an intensity enhancement at the grain boundary. However, these features are emphasized in the NF spectra, which gives the real magnitude of the spectral modifications. As shown in Figure \ref{fig1}f, it can be observed that the grain middle region present a PL spectrum composed by trion and exciton peaks with similar intensities. At the grain boundary, the exciton peak presents a noticeable enhancement while the trion peak is quenched, which is responsible for the observed PL energy blueshift. The exciton/trion peak intensity ratio map shown in Supporting Information Figure S2 reveals a relative enhancement up to 9 times of the exciton intensity with respect to the trion at the grain boundary. This exciton emission increase due to a suppressed trion formation was previously associated to an electrical doping effect \cite{Mouri2013}, which can be locally probed by these near-field measurements \cite{Gadelha2021,Nadas2023}. In fact, a commonly reported feature at MoS$_{2}$ monolayer grain boundaries is the presence of electrical doping effects at these regions \cite{VanderZande2013,Bao2015,Karvonen2017}, related with defects such as sulfur vacancies \cite{Bao2015} or chemical doping \cite{Karvonen2017}. Thus, our results shed light in the nanoscale aspect of the electrical and optical properties modifications of the mismatched grain boundary regions.

\begin{figure}[!htb]
 \centering
 \includegraphics[scale=0.6]{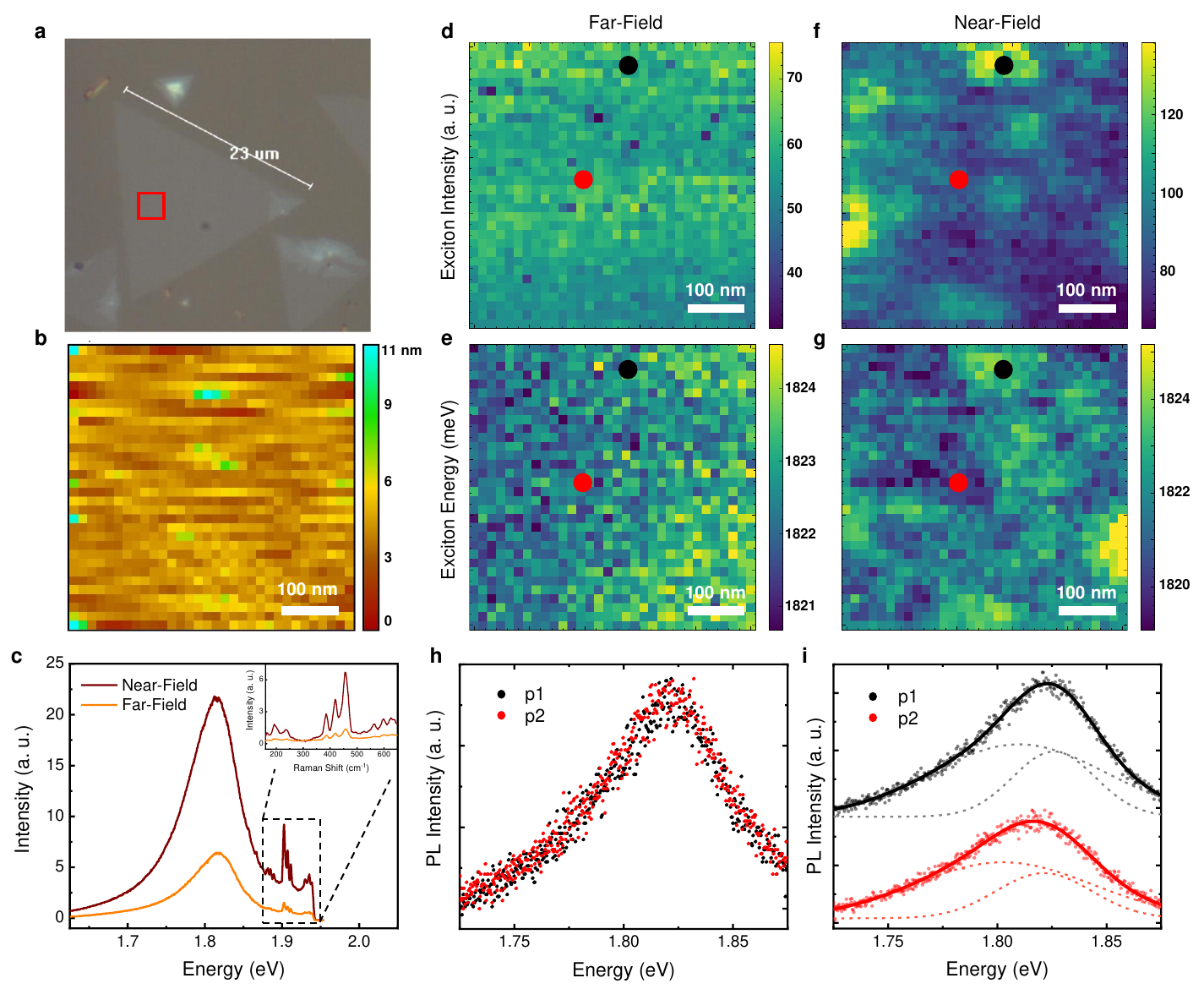} 
 \caption{{\small {\bf a} Optical image of a triangular shaped MoS$_{2}$ monolayer from sample A2. {\bf b} AFM measurement of the $500$x$500$ nm red squared region shown in ({\bf a}). {\bf c} NF and FF PL and Raman spectra of the MoS$_{2}$ monolayer showing the intensity enhancement of the NF measurement. {\bf d-g} PL hyperspectral maps of the FF exciton intensity ({\bf d}), FF exciton energy ({\bf e}), NF exciton intensity ({\bf f}) and NF exciton energy ({\bf g}) in the same red squared region shown in ({\bf a}). The NF maps reveal nanoscale features associated with localized strain fields. The hyperspectral measurements were performed with steps of 16 nm. FF ({\bf h}) and NF ({\bf i}) PL spectra of p1 and p2 points shown by black and red circles in the TEPL hyperspectral maps ({\bf d-g}). The measured spectra are displayed in the scattered data, while the two Gaussian peaks fitting is presented in the solid curves. The trion and exciton Gaussian peaks are highlighted in the dashed curves.}}
 \label{fig2}
 \end{figure}

In the following, we performed TEPL and TERS measurements in the A2 sample in order to study further local heterogeneities. Figure \ref{fig2}a shows the optical image of a triangular shaped MoS$_{2}$ monolayer from the A2 sample. The topography of the $500$x$500$ nm squared region in the center of the monolayer highlighted in red in Figure \ref{fig2}a was probed by a non-contact mode AFM measurement (further details in Methods Section), as displayed in Figure \ref{fig2}b showing no large topographical variations along the sample area. Moreover, Figure \ref{fig2}c shows FF and NF spectra with Raman and PL peaks taken in the middle of the MoS$_{2}$ flake. Spectral enhancement factors of $\sim 3.5$ and $\sim 7$ for PL and Raman intensities, respectively, were observed by engaging the tip. Hyperspectral measurements were also performed in the same region measured by AFM (Figure \ref{fig2}b) to investigate local optical features. The Raman hyperspectrum displayed an uniform response throughout this region in both FF and NF measurements, as shown in Supporting Information Figure S3. However, Figures \ref{fig2}d-g reveal nanoscale modifications in the PL emission over this area. While the FF PL hyperspectrum showed uniform exciton peak intensity (Figure \ref{fig2}d) and energy (Figure \ref{fig2}e) maps, the NF PL hyperspectrum presented local islands of an enhanced exciton intensity (Figure \ref{fig2}f) that matches to an exciton energy blueshift (Figure \ref{fig2}g). To better visualize these features, PL spectra of two distinct positions - p1 (black dot) and p2 (red dot) - are displayed in Figures \ref{fig2}h,i for both FF and NF measurements, respectively. The p1 and p2 position are shown in the PL hyperspectral maps and are related to an enhanced PL signal spot and to a decreased PL signal spot, respectively. Figure \ref{fig2}h presents similar FF PL spectra for p1 and p2, as observed in FF hyperspectral maps. On the other hand, Figure \ref{fig2}i evidences that the NF PL spectra shows an intensity enhancement and energy blueshift at the p1 position. Furthermore, Figure \ref{fig2}i also shows the fitted PL spectra by two Gaussian peaks, respective to the exciton - higher energy peak - and trion - lower energy peak. Supporting Information Figure S4 displays the FF and NF PL maps for the trion emission, presenting a similar behavior with respect to the exciton. It is worth mentioning that all spectra of both FF and NF hyperspectral measurements were fitted by two Gaussian peaks and the hyperspectral maps were made with these fitted parameters. 

As commented, PL intensity enhancements and energy shifts in MoS$_{2}$ monolayers can be associated with doping effects. However, as shown in Supporting Information Figure S5, the exciton/trion peak intensity ratio along the measured area is approximately uniform (variations lower than 35\%), which is not consistent with a doping hypothesis. Another common feature reported in TMDs that induces PL energy blueshift and intensity enhancement is the variation of strain fields \cite{Liu2014,Castellanos-Gomez2015,Christopher2019,Li2020} along the sample. CVD grown MoS$_{2}$ monolayers have previously presented shifts around $40$ meV/$\%$ of strain for the exciton emission \cite{Liu2014,Christopher2019}. As we observe shifts up to $\sim 4$ meV for the exciton peak, it would be associated with a strain of $\sim 0.1\%$. Indeed, this strain is not sufficient to be probed by Raman, as it would induce shifts of $\sim 0.21$ cm$^{-1}$ and $\sim 0.07$ cm$^{-1}$ in the E$_{2g}$ and A$_{1g}$ peaks \cite{Christopher2019}, respectively, that is below our spectrometer resolution. Moreover, the PL intensity enhancement is also in accordance with the $0.1\%$ strain. Besides, as there is no coincidence between the AFM measurement and the local variation presented in the NF PL maps, the suggestive strain is not accounted on a topographic reason. As shown in Supporting Information Figure S6, this strained regions are also noticed in another region of the flake. Therefore, the observed non-uniform localized strain is presumably due to an expected thermal expansion coefficient and lattice constant mismatch between the monolayer and the substrate \cite{Deng2018}, which plays a major role in the growth process. Although strain is a well studied characteristic of TMDs, unveiling strained responses at this nanoscale can raise the understanding of the substrate and growth method roles in the optical properties of these 2D semiconductors.

\begin{figure}[!htb]
 \centering
 \includegraphics[scale=0.5]{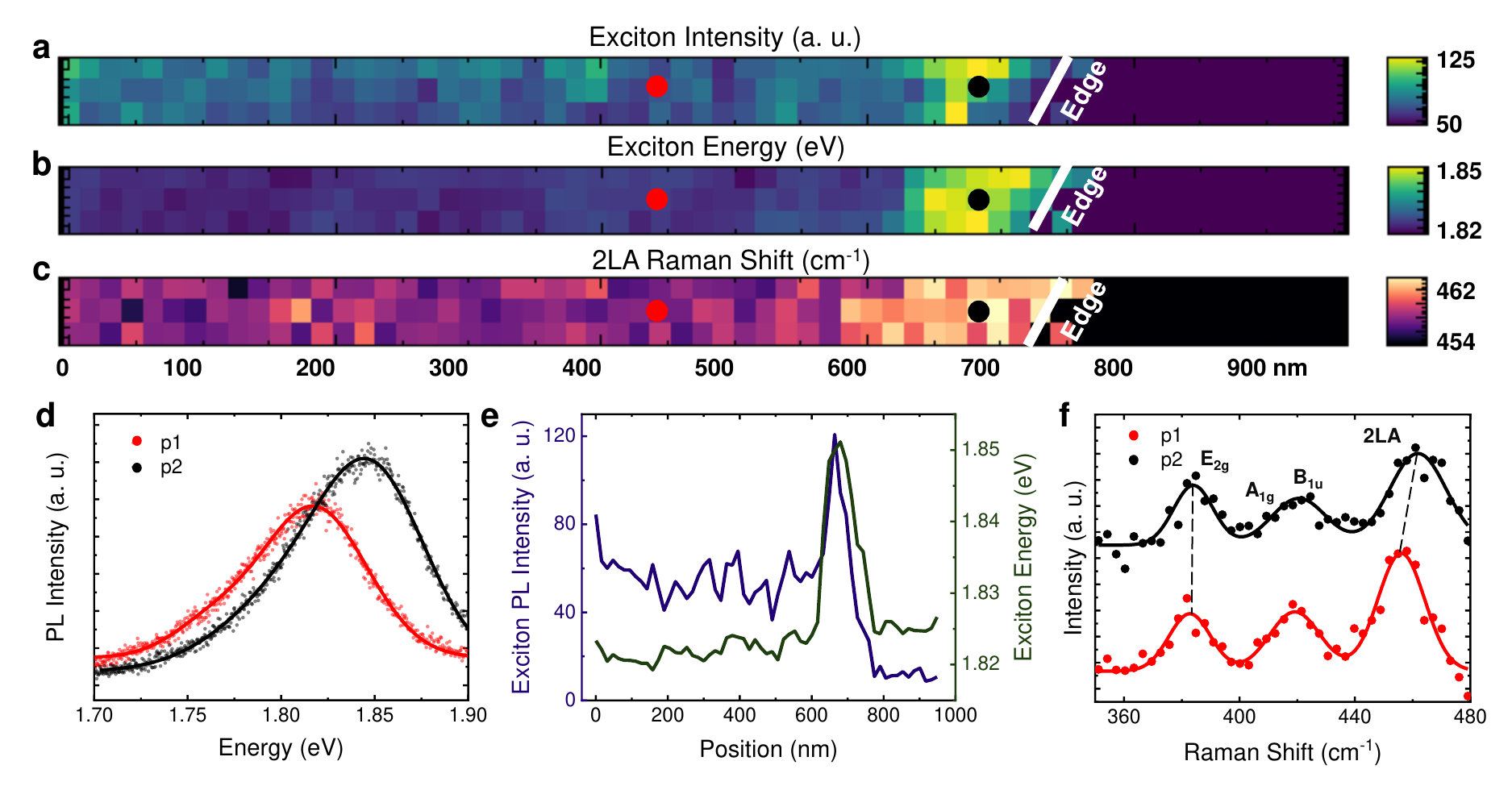} 
 \caption{{\small {\bf a,b} NF TEPL hyperspectral maps of the exciton intensity ({\bf a}) and energy ({\bf b}) at an edge region of the MoS$_{2}$ monolayer shown in Figure \ref{fig2}. An intensity enhancement and an energy blueshift of the PL is observed at the edges. {\bf c} NF TERS hyperspectral frequency map of the 2LA mode, revealing a blueshift of $\sim 4$ cm$^{-1}$ at the edge. The PL and Raman hyperspectral measurements were performed with steps of 16 nm. {\bf d} NF PL spectra of p1 and p2 points shown by black and red circles in the hyperspectral maps ({\bf a-c}). The measured spectra are displayed in the scattered data, while the spectra fitting is presented in the solid curves. {\bf e} Exciton PL intensity and energy profiles along the edge, highlighting its enhancement and blueshift of $\sim 25$ meV. Moreover, the intensity profile reveal a spatial resolution of $\sim 30$ nm. {\bf f} NF Raman spectra of p1 and p2 points shown by black and red circles in the hyperspectral maps ({\bf a-c}). The measured spectra are displayed in the scattered data, while the spectra fitting is presented in the solid curves.}}
 \label{fig3}
 \end{figure}

Another frequent heterogeneity reported in monolayer TMDs is associated to their edges \cite{Lin2016}. Thereby, we have also performed TEPL and TERS hyperspectral measurements in an edge region of the same MoS$_{2}$ flake shown in Figure \ref{fig2}. Figures \ref{fig3}a,b display the exciton peak intensity and energy maps for the NF PL hyperspectral measurement in this edge region. An intensity enhancement and an energy blueshift of the exciton peak are noted at the edge within a spatial width of less than 100 nm, which was not observed in the FF measurements as shown in Supporting Information Figure S7. These features are also presented in Figure \ref{fig3}d, in which the PL spectra taken at the middle (p1) and at the edge (p2) of the sample confirm the intensity enhancement and energy shift at the edge. Moreover, Supporting Information Figure S8 also shows similar results for another edge region of the same sample. Unlike the PL energy shift in the inferred strained regions throughout the monolayer (Fig. \ref{fig2}), the observed energy shift at the edges are remarkably larger. As can be noticed in the exciton energy profile in Figure \ref{fig3}e, there is an energy shift of $25$ meV at the edge. Moreover, this blueshift corresponds to a PL enhancement of $\sim 1.5$, as can also be seen in Figure \ref{fig3}d. If this observed PL energy shift was also due to strain, it would be associated with measurable Raman shifts. However, a NF Raman hyperspectral measurement showed no significant shift for the E$_{2g}$ mode at the edge as shown in Supporting Information Figure S9. On the other hand, the 2LA Raman mode presented a noticeable blueshift of $\sim 4$ cm$^{-1}$ as displayed in the TERS peak position map (Figure \ref{fig3}c) and its profile along the edge (Supporting Information Figure S9). The absence of frequency shifts of the first order Raman modes as well as the blueshift of the 2LA mode is also shown in the Raman spectra of the middle p1 position and at the edge p2 position shown in Figure \ref{fig3}f. As shown by Carvalho \textit{et al.} \cite{Carvalho2017}, the MoS$_{2}$ monolayer 2LA Raman mode is in fact composed by different Raman peaks that are highly dependent with electronic resonances. As these peaks present distinct resonances by varying the laser wavelength \cite{Carvalho2017}, a similar behavior is expected if the material band gap is modified. Therefore, the observed 2LA frequency shift at the edge is presumably related to a modification in the relative intensities of the 2LA peaks due to the band gap shift noticed in the PL measurements. Previous reports \cite{Birmingham2018,Kumar2018,Hu2019,Rodriguez2019} have observed different PL features at MoS$_{2}$ monolayer edges due to distinct defects. Water and oxygen passivation of vacancies at the edges also result in an enhancement in the PL intensity, but it corresponds to a redshift in the PL energy \cite{Birmingham2018,Hu2019}. On the other hand, a similar nanoscale edge response was recently probed by hyperspectral TEPL measurements \cite{Rodriguez2019}. However, an assertive determination of the defect responsible for these optical modifications at the edges is still lacking. Therefore, the association of this local PL enhancement and blueshift at the edges with a frequency shift of the 2LA mode brings more information of the influence of the edge defective states in the MoS$_{2}$ monolayer optical properties.

\begin{figure}[!htb]
 \centering
 \includegraphics[scale=0.55]{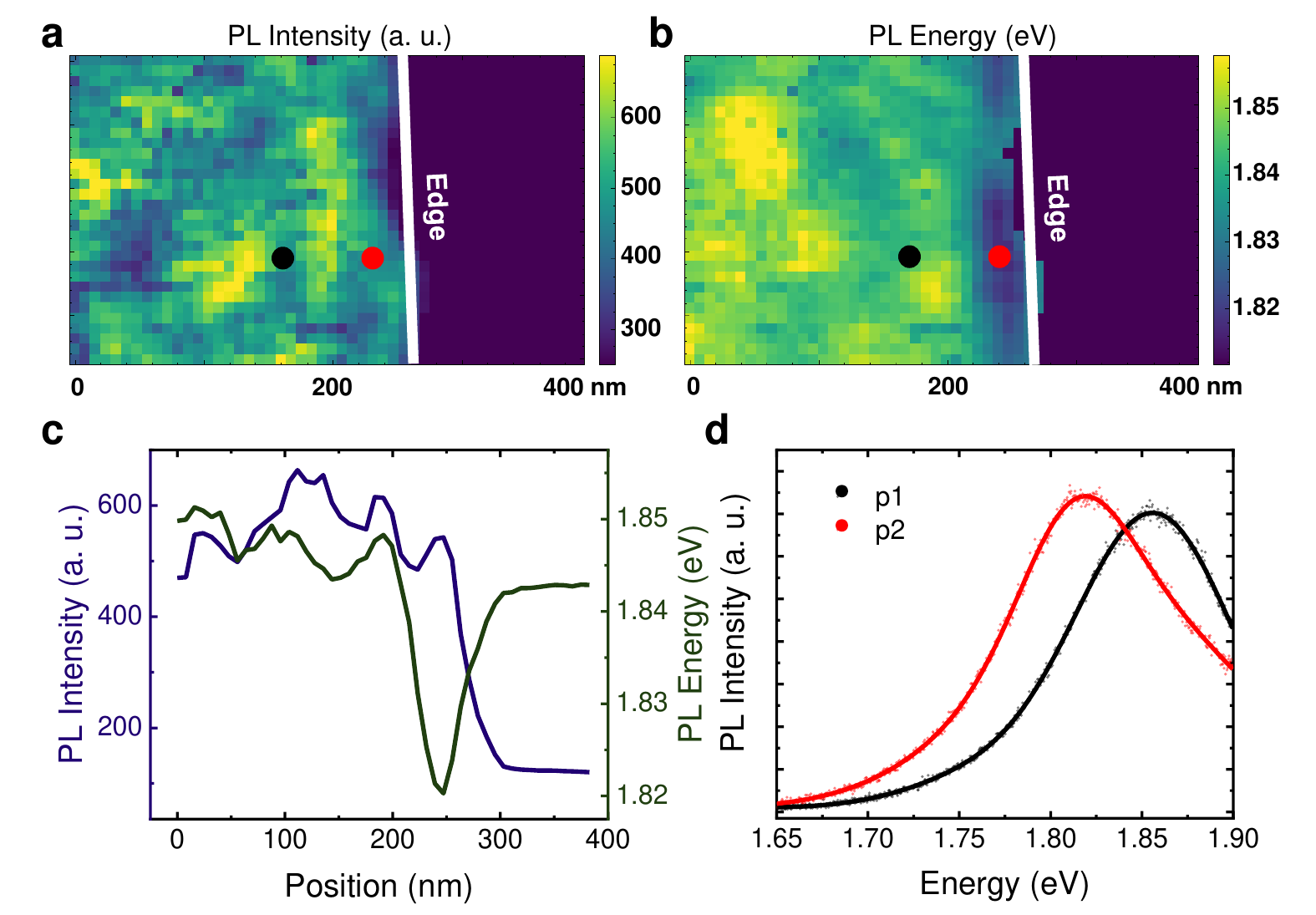} 
 \caption{{\small {\bf a,b} NF TEPL hyperspectral maps of the PL intensity ({\bf a}) and energy ({\bf b}) at an edge region of a MoS$_{2}$ monolayer from A1 sample. An energy redshift of the PL is observed in at the edges. The hyperspectral measurements were performed with steps of 8 nm. {\bf c} Exciton PL intensity and energy profiles along the edge, highlighting its redshift of $\sim 25$ meV. Moreover, the intensity profile reveal a spatial resolution of $\sim 20$ nm. {\bf c} NF PL spectra of p1 and p2 points shown by black and red circles in the TEPL hyperspectral maps ({\bf a,b}). The measured spectra are displayed in the scattered data, while the spectra fitting is presented in the solid curves. The trion and exciton Gaussian peaks are highlighted in the dashed curves for the p2 point, showing that the observed redshift in the PL energy is due to a predominant trion formation.}}
 \label{fig4}
 \end{figure}

In order to further investigate other types of optical responses at the edges, we have measured other MoS$_{2}$ monolayer edges from the A1 sample. Figures \ref{fig4}a,b shows the measured NF PL hyperspectrum of a A1 MoS$_{2}$ monolayer edge. Conversely to the observed features of sample A2, the PL intensity map of Figure \ref{fig4}a presents no clear modification, whereas the PL energy map of Figure \ref{fig4}b reveals a significant redshift in an edge region of less than $50$ nm of width. TEPL measurements in distinct edge regions of different monolayer flakes of the same A1 sample presented a similar PL energy redshift, as displayed in Supporting Information Figure S10. Figure \ref{fig4}c shows PL intensity and energy profiles along the edge, in which the energy redshift of $25$ meV can be observed. Moreover, in the intensity profile is possible to extract the spatial resolution of the measurement, that is around $20$ nm (see further details in Supporting Information Figure S11). PL spectra at the middle (p1 position) and at the edge (p2 position) of the sample are shown in Figure \ref{fig4}d, clearly showing the PL redshift presented in the TEPL hiperspectral map and profile, but no clear spectral modification. Although there is a PL energy redshift associated with oxygen and water passivation at the edges \cite{Birmingham2018,Hu2019}, the absence of a strong PL intensity enhancement suggests a different reason for this observed edge response. On the other hand, if this intense PL energy modification at the edge is related with strain, a shift in the first order Raman modes should be observed.

\begin{figure}[!htb]
 \centering
 \includegraphics[scale=0.36]{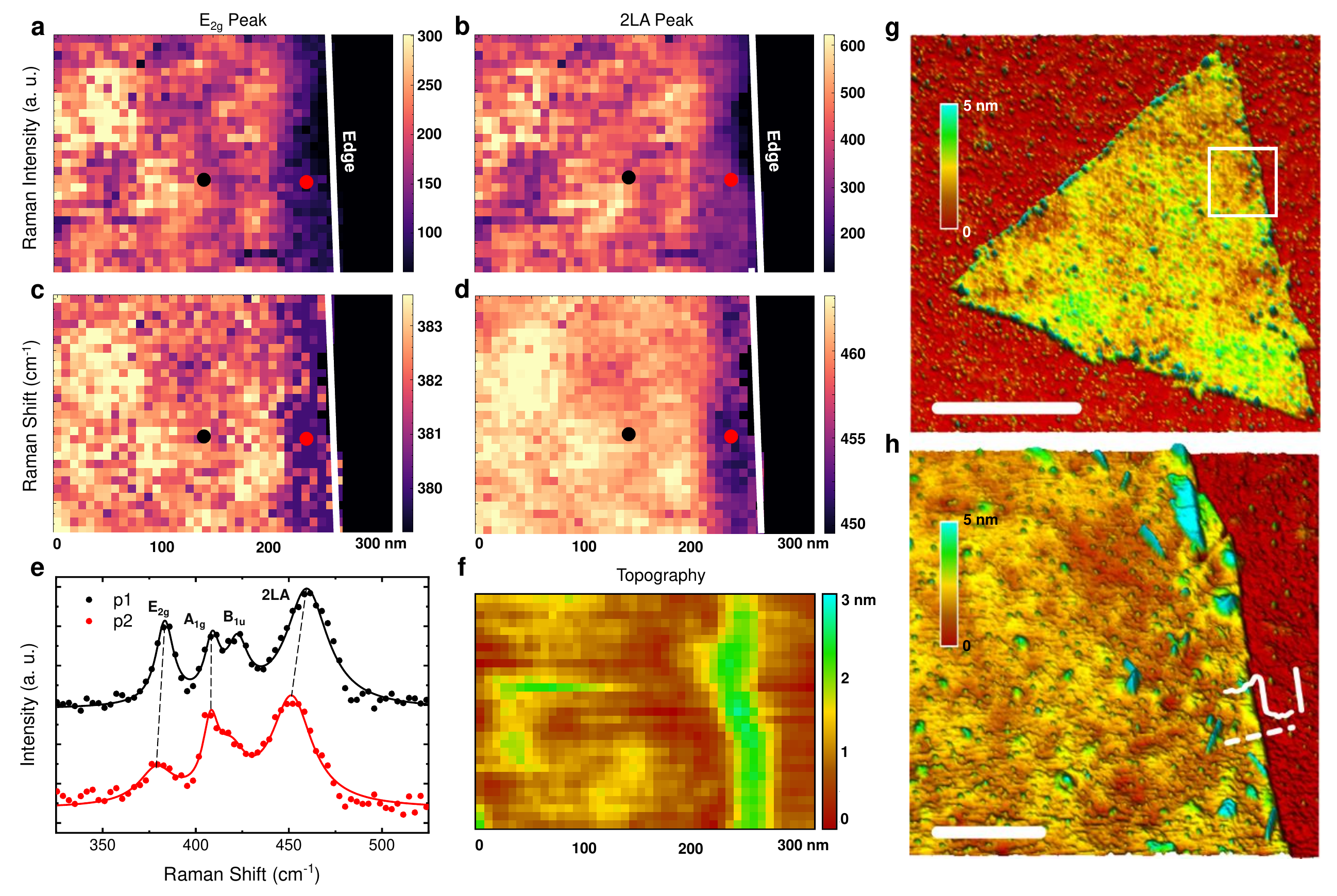} 
 \caption{{\small {\bf a-d} NF TERS hyperspectral maps of the E$_{2g}$ Raman intensity ({\bf a}) and Raman shift ({\bf c}), and 2LA Raman intensity ({\bf b}) and Raman shift ({\bf d}) for a MoS$_{2}$ monolayer. The TERS hyperspectral measurements were performed with steps of 8 nm and in the same region of the TEPL hyperspectral measurements of Figure \ref{fig4}. They reveal a remarkable frequency redshift for both E$_{2g}$ and 2LA Raman modes, that is highlighted in NF Raman spectra of p1 and p2 points ({\bf e}). These points are shown by black and red circles in ({\bf a-d}) maps and are related to an inner and an edge spot of the sample, respectively. {\bf f} Topography measurement in the same region of the hyperspectral maps using the TERS tip height. The edge shows a noticeable suspension that is directly associated with a strain field. {\bf g} AFM measurement of the whole monolayer showing the extended edge suspension. {\bf h} AFM measurement of the region underlined in the inset white square in ({\bf h}) revealing the edge topographic response in this focused area. The inset height profile in ({\bf h}) confirms the edge suspension. Scale bar in ({\bf g}): 5 $\mu$m. Scale bar in ({\bf h}): 500 nm. Scale bar of the height profile in ({\bf h}): 3 nm.}}
 \label{fig5}
 \end{figure}

 Therefore, we also performed a NF Raman hyperspectral measurement in the same region as the measurements of Figure \ref{fig4}. The Raman intensity maps of the E$_{2g}$ and 2LA modes are displayed in Figures \ref{fig5}a,b, while the frequency maps of these peaks are presented in Figures \ref{fig5}c,d. For both peaks it is observed an intensity quenching and a frequency redshift at the edge. Figure \ref{fig5}e shows Raman spectra taken in the middle of the monolayer (p1) and at its edge (p2), highlighting the features noted in the Raman maps. To better quantify the frequency shifts of the Raman modes, their intensity and frequency profiles along the edge are shown in Supporting Information Figure S12. It can be observed a $-1.5$ cm$^{-1}$ shift at the edge for the E$_{2g}$ peak and a $-8.5$ cm$^{-1}$ for the 2LA mode. As commented for the A2 sample, this  2LA shift can also be associated with the modification in the relative intensities of the 2LA peaks due to their distinct resonant responses \cite{Carvalho2017}. Moreover, it is worth commenting that the PL blueshift at the edge of A2 sample related to a 2LA frequency blueshift is in accordance with the PL redshift of the A1 sample at its edge corresponding to a 2LA redshift. Besides, for a strain hypothesis, the $1.5$ cm$^{-1}$ redshift of the E$_{2g}$ mode is in agreement with the observed PL redshift of $25$ meV \cite{Christopher2019}, which would be associated with a strain of approximately $0.7\%$. Another commonly reported feature associated with strain with this magnitude in MoS$_{2}$ monolayer is the splitting of the E$_{2g}$ mode in two peaks due to a lattice symmetry breaking \cite{Conley2013}. Although we did not have the spectral resolution to resolve these two peaks, the broadening of the E$_{2g}$ mode observed in the Raman spectrum at the edge of Figure \ref{fig5}e is another indication of strain. Furthermore, the Raman intensity and frequency maps and profiles of the A$_{1g}$ mode are shown in Supporting Information Figure S13. Whereas there is an intensity enhancement at the edge, no substantial shift is noted for the A$_{1g}$ peak. The expected A$_{1g}$ shift for a $0.7\%$ strain is $-0.5$ cm$^{-1}$, which was probably not detected due to our measurement spectral resolution and its convolution with the B$_{1u}$ mode.

 In order to further corroborate the strain field responsible for the nanoscale optical features at the edge we performed AFM measurements in the A1 sample to probe possible topographic disorders. Simultaneously with the acquisition of the TEPL and TERS measurements, the topographic response of the tip (non-contact mode AFM) was also measured as shown in Figure \ref{fig5}f. A noticeable suspension in the edge region of less than $50$ nm of width can be observed, indicating the presence of a strain field there. In order to confirm this result, high resolution AFM images was measured as displayed in Figures \ref{fig5}g,h. Figure \ref{fig5}g present the AFM for the whole monolayer flake, while Figure \ref{fig5}h shows the AFM for the edge region highlighted in the inset white square of Figure \ref{fig5}g. Besides the surface roughness across the monolayer, both measurements presented a distinct topographic response at the edge, underlined by the height profile in the inset of Figure \ref{fig5}h.  Thus, the sample indeed shows a localized topographic variation at the edges that can generate the observed strain fields that could be only probed by NF PL and Raman measurements.

 In summary, we studied local defects in distinct CVD grown MoS$_{2}$ monolayers  by NF optical measurements. Electronic structure modifications in nanoscale regions along MoS$_{2}$ monolayer grain boundaries were probed by TEPL, in which a remarkable intensity enhancement of the exciton peak related to a quenching in the trion peak emission due to a doping effect was revealed. Besides, local strain fields were observed throughout a MoS$_{2}$ monolayer by TEPL and TERS measurements, revealing the role of the lattice mismatch between sample and substrate. Finally, edge defects were investigated by TEPL and TERS in MoS$_{2}$ monolayers. While one of the grown MoS$_{2}$ monolayer presented an PL intensity enhancement and energy blueshift at the edges, the other grown sample showed a PL energy redshift in the edge regions. Moreover, this last sample displayed a frequency redshift in the E$_{2g}$ and 2LA Raman modes as well as a topographic suspension revealed by the AFM measurement, which suggest a strong localized strain field at the edges. Our work highlights the importance of utilizing optical spectroscopies with nanometric resolution to probe localized defects and reveal the optical properties in two-dimensional materials, that would be otherwise hindered in typical micrometer resolved spectroscopies.

 \section{Methods}

\subsection{Sample Preparation}

Monolayers A1 and A2 of MoS$_2$ were synthesized on fused silica substrates
using the CVD technique. Two different single-zone
quartz tube furnaces were employed for the growth process, utilizing MoO$_3$
($>99.5\%$ purity, Aldrich) and sulfur powder (Aldrich, $>99\%$) as precursors.
Substrates were subjected to ultrasonic cleaning in baths of acetone and
isopropyl alcohol for 10 minutes each.

For A1, the monolayer was grown at atmospheric pressure inside a 1-inch
diameter quartz tube. The cleaned substrate was positioned over a crucible
containing 1.1\,mg of MoO$_3$, centered in the tube coinciding with the position of the central thermopar. Another crucible, situated approximately 17\,cm upstream from the substrate, held 234 mg of sulfur powder. Ar gas purging was carried out for 20 minutes at a flow rate of 180 standard cubic centimeters per minute (sccm), and during the growth process, a 40
sccm Ar gas flow served as the carrier. The sample temperature underwent an
initial ramp to 300 °C at a rate of 7.5 °C.min$^{-1}$, followed by a second ramp from
300 to 850 °C at a rate of 15 °C.min$^{-1}$, with a 5-minute hold at the final
temperature. At a central temperature of 600 °C, sulfur vapor was introduced by
ramping up the temperature of the thermal rubber band at a rate of 50 °C.min$^{-1}$
until reaching 150 °C, a temperature maintained throughout the growth until
complete sulfur evaporation. Following growth, the furnace was powered off and
allowed to cool naturally.

For A2, the monolayer was grown under atmospheric pressure inside a 2.5-inch
diameter quartz tube. In this case, the cleaned fused silica substrate was placed
over a cleaned Si/SiO$_2$ (300nm) substrate. The stacked substrates were
positioned over a crucible containing 1.1 mg of MoO$_3$, centered in the tube
coinciding with the central thermopar. Another crucible, situated approximately
21 cm upstream from the substrate, held 350 mg of sulfur powder. Ar gas purging
and carrier gas conditions were identical to those of A1. The sample temperature
underwent an initial ramp to 300 °C at a rate of 7.5 °C.min$^{-1}$, followed by a
second ramp from 300 to 800 °C at a rate of 15 °C.min$^{-1}$, with a 15-minute hold at the final temperature. Sulfur vaporization commenced when the central
temperature reached 600 °C. Following growth, the furnace was powered off and
allowed to cool naturally.

\subsection{Optical Measurements}

In our TERS and TEPL measurements, we employed the FabNS Porto-SNOM system, utilizing an oil-immersion objective lens (NA=1.4), in conjunction with a radially polarized HeNe laser (632.8 nm). The probe tips utilized were the Plasmon-Tunable Tip Pyramid (PTTP) \cite{vasconcelos2018plasmon}, tuned to match the laser wavelength. The acquired spectral data were subsequently analyzed with the PortoFlow Analysis Software (v1.15).

The SHG measurements were carried out by using a picosecond OPO laser (APE Picoemerald), with an incident power of 50 mW at the sample and a 810 nm excitation wavelength. The SHG imaging was performed by scanning the laser with a set of galvanometric mirrors (LaVison BioTec) in a Nikon microscope. The laser beam was focused on the sample by a 40x objective with numerical aperture N.A. = 0.95 after a half-wave plate and the backscattered signal was collected by the same objective and directed to a photomultiplier tube (PMT). We placed an analyzer in front of the PMT that we rotated together with the half-wave plate, as well as a 405/10 nm band-pass filter to collect only the SHG signal. The SHG images were taken in steps of 2° of the half wave plate and analyzer.

\subsection{AFM Measurements}

High resolution AFM experiments were carried out using a XE-70 atomic force microscope from Park Systems, Korea. All images were acquired in the intermittent contact mode and in ambient conditions. Si cantilevers (DPE/XSC11 hard, from MikroMasch), with spring constants of 3-16 N.m$^{-1}$ and a tip radius of curvature $\sim$10 nm, were used throughout the study for sample imaging. These AFM images were processed (leveling, profiling, and 3D rendering) using the open-source software package Gwyddion.

The topographic response of the tip on the TERS and TEPL measurements were done by a shear force detection feedback AFM (non-contact mode).

\begin{acknowledgement}

The authors acknowledge financial support from CNPq, CAPES, FAPEMIG, FINEP, Brazilian Institute of Science and Technology (INCT) in Carbon Nanomaterials and Rede Mineira de Materiais 2D (FAPEMIG).

\end{acknowledgement}

\begin{suppinfo}

Polarized SHG. Exciton/trion intensity ratio maps. Far-field and near-field Raman maps. Far-field and near-field trion maps. Far-field and near-field exciton maps. Near-field spatial resolution. Near-field Raman modes intensity and frequency profiles.

\end{suppinfo}


\providecommand{\latin}[1]{#1}
\makeatletter
\providecommand{\doi}
  {\begingroup\let\do\@makeother\dospecials
  \catcode`\{=1 \catcode`\}=2 \doi@aux}
\providecommand{\doi@aux}[1]{\endgroup\texttt{#1}}
\makeatother
\providecommand*\mcitethebibliography{\thebibliography}
\csname @ifundefined\endcsname{endmcitethebibliography}
  {\let\endmcitethebibliography\endthebibliography}{}

\newpage

\title{}
{\noindent\large{\textbf {%
\centering{%
Supporting Information for \\
Nano-optical investigation of grain boundaries, strain and edges in CVD grown MoS$_{2}$ monolayers
}}}
}
\\

Frederico B. Sousa, Rafael Battistella Nadas, Rafael Martins, Ana P. M. Barboza, Bernardo R. A. Neves, Ive Silvestre, Ado Jorio, and Leandro M. Malard.

%

\paragraph*{This Supporting Information includes:\newline}
   %

    $\bullet$ Figure~S1.  Optical image and polarized SHG of MoS$_2$ monolayer sample A1. \\
    $\bullet$ Figure~S2. PL maximum energy and exciton/trion intensity ratio maps at the grain boundary of MoS$_2$ monolayer sample A1. \\
    $\bullet$ Figure~S3. Far-field and near-field Raman maps of MoS$_2$ monolayer sample A2. \\
    $\bullet$ Figure~S4. Far-field and near-field trion maps of MoS$_2$ monolayer sample A2. \\
    $\bullet$ Figure~S5. Exciton/trion intensity ratio map of MoS$_2$ monolayer sample A2. \\
    $\bullet$ Figure~S6. Near-field exciton maps of MoS$_2$ monolayer sample A2. \\
    $\bullet$ Figure~S7. Far-field exciton maps at the edge of MoS$_2$ monolayer sample A2. \\
    $\bullet$ Figure~S8. Near-field exciton maps at the edge of MoS$_2$ monolayer sample A2. \\
    $\bullet$ Figure~S9. Near-field Raman maps at the edge of MoS$_2$ monolayer sample A2. \\
    $\bullet$ Figure~S10. Near-field exciton maps at the edge of MoS$_2$ monolayer sample A1. \\
    $\bullet$ Figure~S11. Near-field spatial resolution. \\
    $\bullet$ Figure~S12. Near-field Raman modes intensity and frequency profiles along the edge of MoS$_2$ monolayer sample A1. \\
    $\bullet$ Figure~S13. Near-field A$_{1g}$ maps at the edge of MoS$_2$ monolayer sample A1. \\


\newpage

\begin{figure}[!htb]
 \centering
 \includegraphics[scale=0.45]{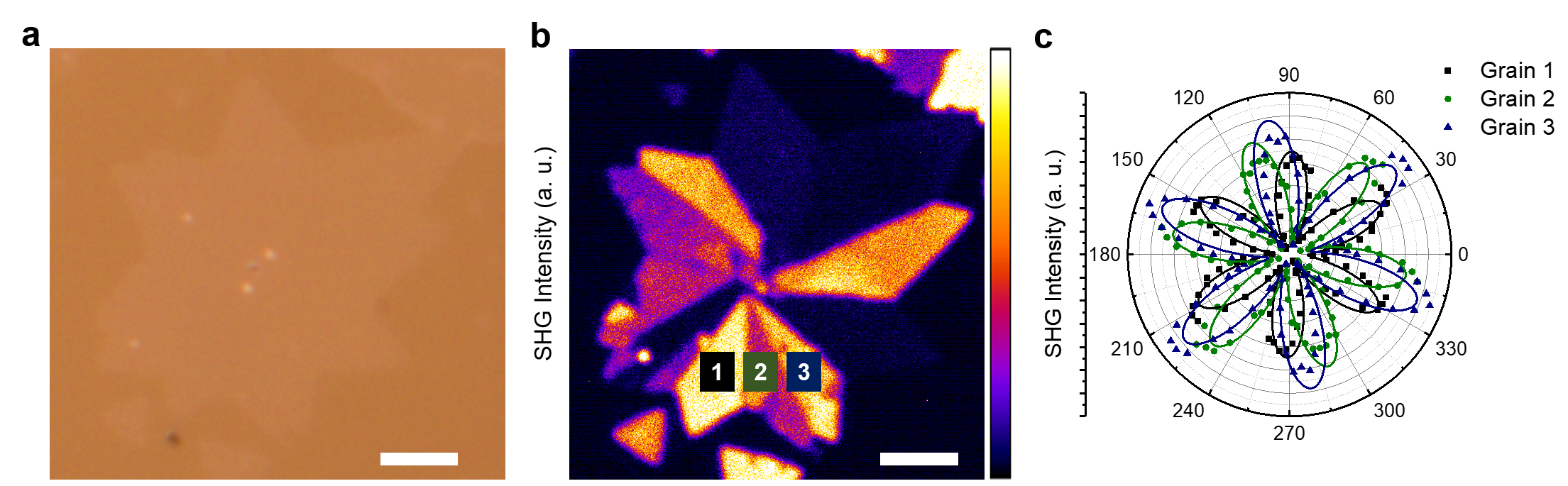} 
 \caption{\small {\bf a} Optical image of MoS$_{2}$ monolayer sample A1 shown in Figure 1. {\bf b} SHG intensity image of MoS$_{2}$ monolayer sample A1 with the 3 studied grains highlighted. Scale bars: 5 $\mu$m. {\bf c} Polarized SHG measurement for grains 1, 2 and 3 of MoS$_{2}$ monolayer sample A1. The relative orientations between them are: $\theta_{1,2}$ = 21°, $\theta_{1,3}$ = 11° and $\theta_{2,3}$ = 10°.}
 \label{SI_Fig1}
 \end{figure}


\begin{figure}[!htb]
 \centering
 \includegraphics[scale=0.6]{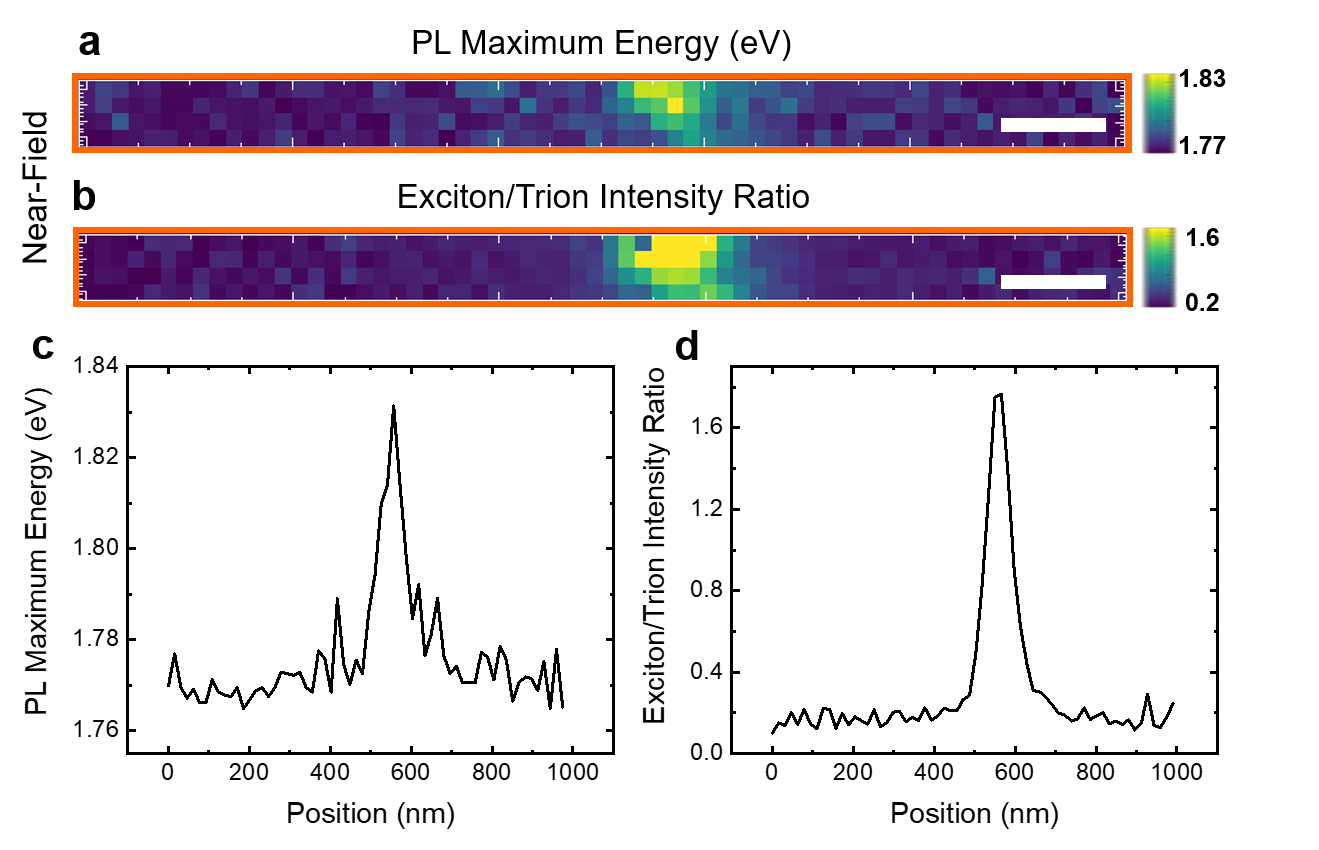} 
 \caption{\small {\bf a,b} Near-field PL maximum energy and exciton/trion intensity ratio maps at a grain boundary of MoS$_{2}$ monolayer sample A1. These maps are from the orange rectangle region highlighted in Figure 1b. Scale bars: 50 nm. {\bf c,d} PL maximum energy and exciton/trion intensity ratio profiles along the grain boundary of MoS$_{2}$ monolayer sample A1. }
 \label{SI_Fig2}
 \end{figure}


 \begin{figure}[!htb]
 \centering
 \includegraphics[scale=0.45]{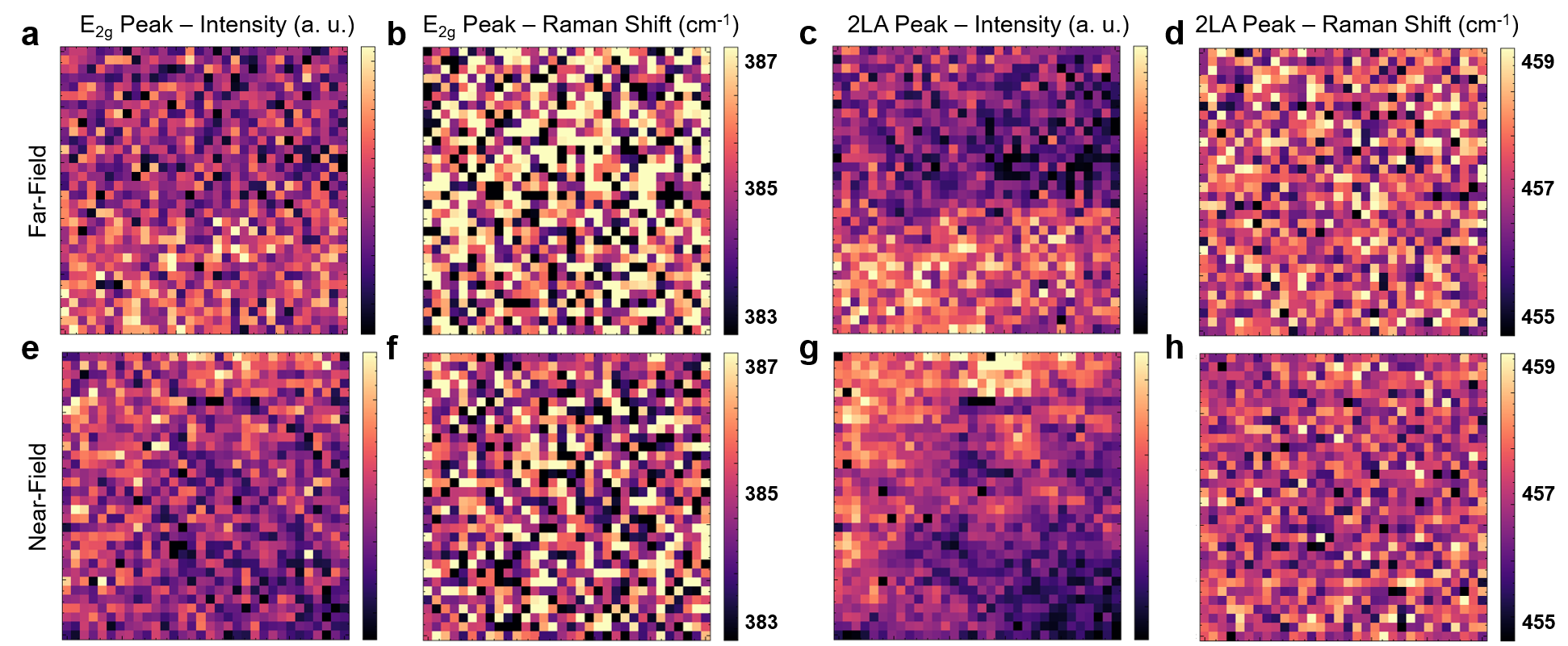} 
 \caption{\small {\bf a-d} Far-field E$_{2g}$ intensity ({\bf a}), E$_{2g}$ frequency ({\bf b}), 2LA intensity ({\bf c}), and 2LA frequency ({\bf d}) maps of MoS$_{2}$ monolayer sample A2. {\bf e-h} Near-field E$_{2g}$ intensity ({\bf e}), E$_{2g}$ frequency ({\bf f}), 2LA intensity ({\bf g}), and 2LA frequency ({\bf h}) maps of MoS$_{2}$ monolayer sample A2. These maps are from the same region of the exciton maps of Figure 2.}
 \label{SI_Fig3}
 \end{figure}


 \begin{figure}[!htb]
 \centering
 \includegraphics[scale=0.5]{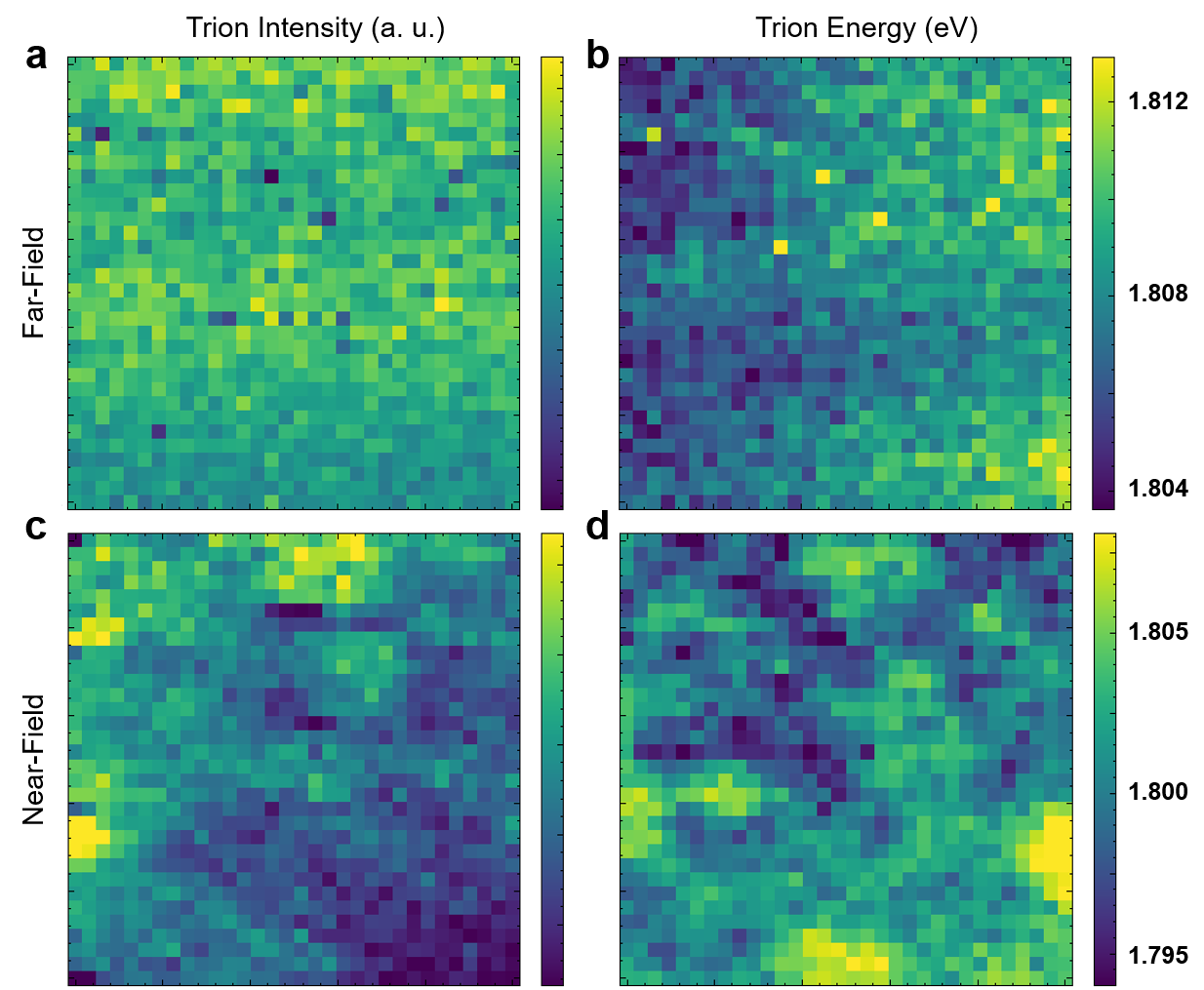} 
 \caption{\small {\bf a,b} Far-field trion intensity ({\bf a}) and trion energy ({\bf b}) maps of MoS$_{2}$ monolayer sample A2. {\bf c,d} Near-field trion intensity ({\bf c}) and trion energy ({\bf d}) maps of MoS$_{2}$ monolayer sample A2. These maps are from the same region of the exciton maps of Figure 2.}
 \label{SI_Fig4}
 \end{figure}


 \begin{figure}[!htb]
 \centering
 \includegraphics[scale=0.5]{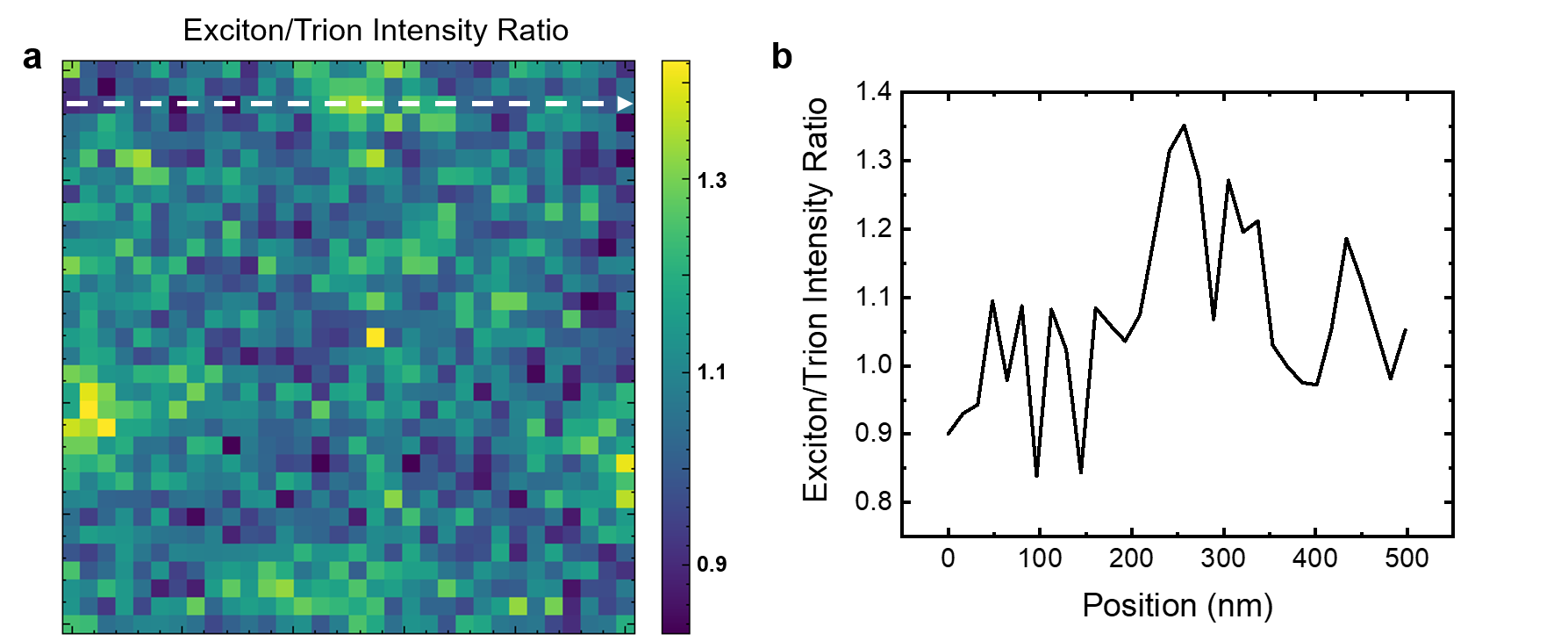} 
 \caption{\small {\bf a} Near-field exciton/trion intensity ratio map of MoS$_{2}$ monolayer sample A2. This map is from the same region of the exciton maps of Figure 2. {\bf b} Exciton/trion intensity ratio profile taken along the dashed white arrow shown in ({\bf a}).}
 \label{SI_Fig5}
 \end{figure}


\begin{figure}[!htb]
 \centering
 \includegraphics[scale=0.7]{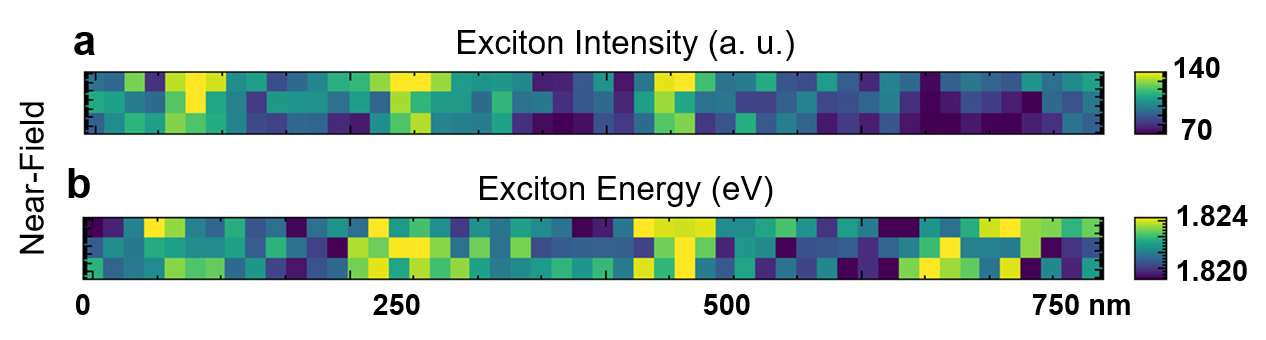} 
 \caption{\small {\bf a,b} Near-field exciton intensity ({\bf a}) and exciton energy ({\bf b}) maps of MoS$_{2}$ monolayer sample A2 showing localized strain fields in a different sample region.}
 \label{SI_Fig6}
 \end{figure}


\begin{figure}[!htb]
 \centering
 \includegraphics[scale=0.7]{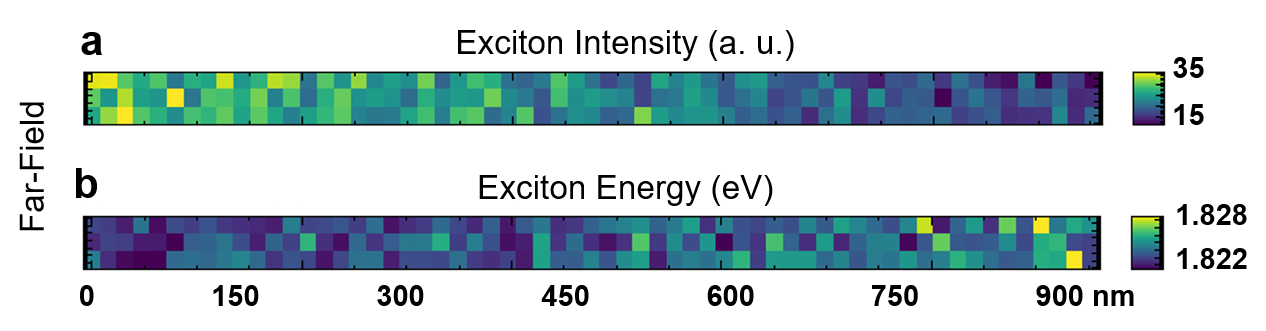} 
 \caption{\small {\bf a,b} Far-field exciton intensity ({\bf a}) and exciton energy ({\bf b}) maps at the edge of MoS$_{2}$ monolayer sample A2 showing no localized optical features. These maps are from the same region of the maps of Figure 3.}
 \label{SI_Fig7}
 \end{figure}


\begin{figure}[!htb]
 \centering
 \includegraphics[scale=0.6]{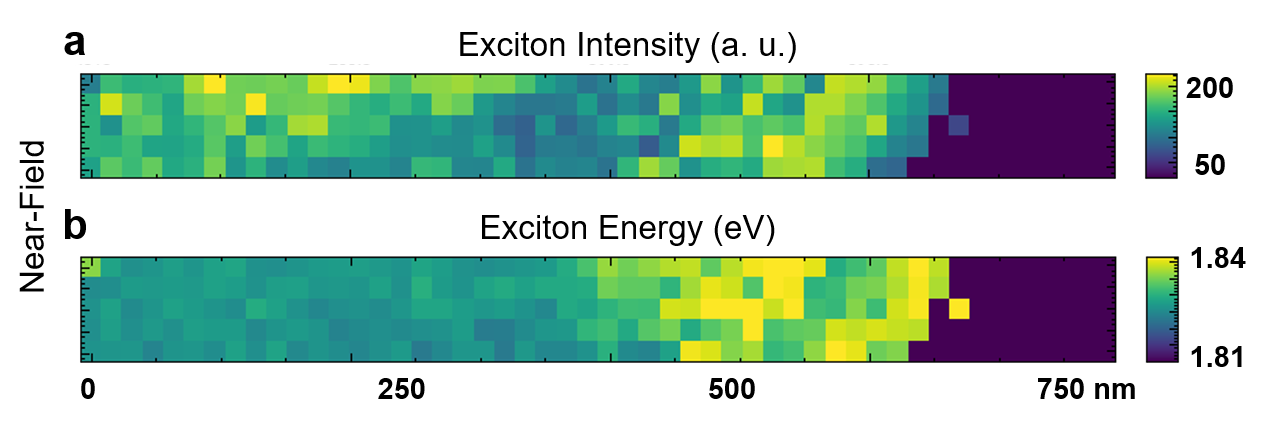} 
 \caption{\small {\bf a,b} Near-field exciton intensity ({\bf a}) and exciton energy ({\bf b}) maps in another edge region of MoS$_{2}$ monolayer sample A2 showing similar PL enhancement and blueshift features presented at the edge region of Figure 3.}
 \label{SI_Fig8}
 \end{figure}


\begin{figure}[!htb]
 \centering
 \includegraphics[scale=0.7]{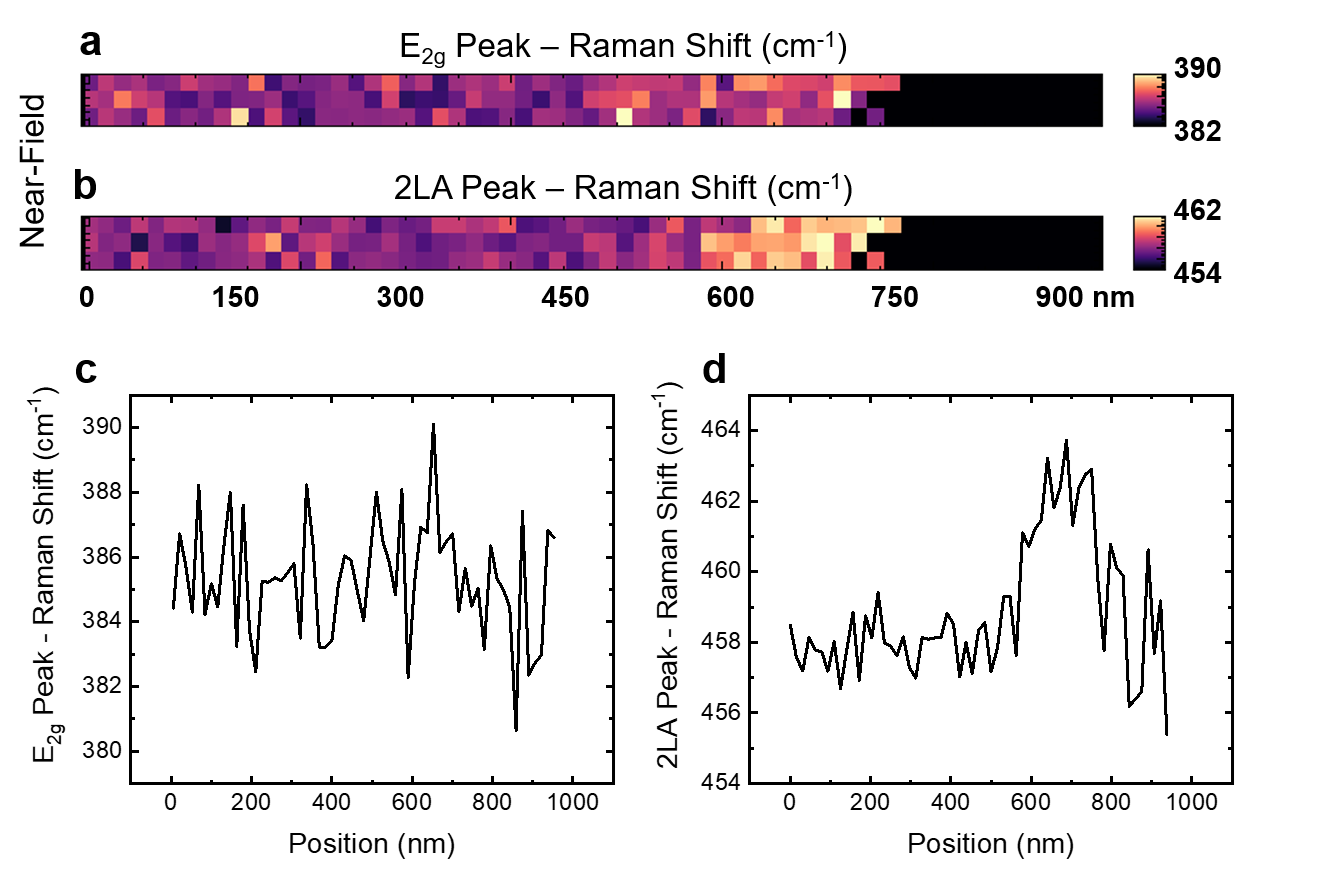} 
 \caption{\small {\bf a,b} Near-field E$_{2g}$ ({\bf a}) and 2LA ({\bf b}) frequency maps at the edge of MoS$_{2}$ monolayer sample A2. These maps are from the same region of the maps of Figure 3. {\bf c,d} E$_{2g}$ ({\bf c}) and 2LA ({\bf d}) frequency profiles along the edge of MoS$_{2}$ monolayer sample A2.}
 \label{SI_Fig9}
 \end{figure}


 \begin{figure}[!htb]
 \centering
 \includegraphics[scale=0.45]{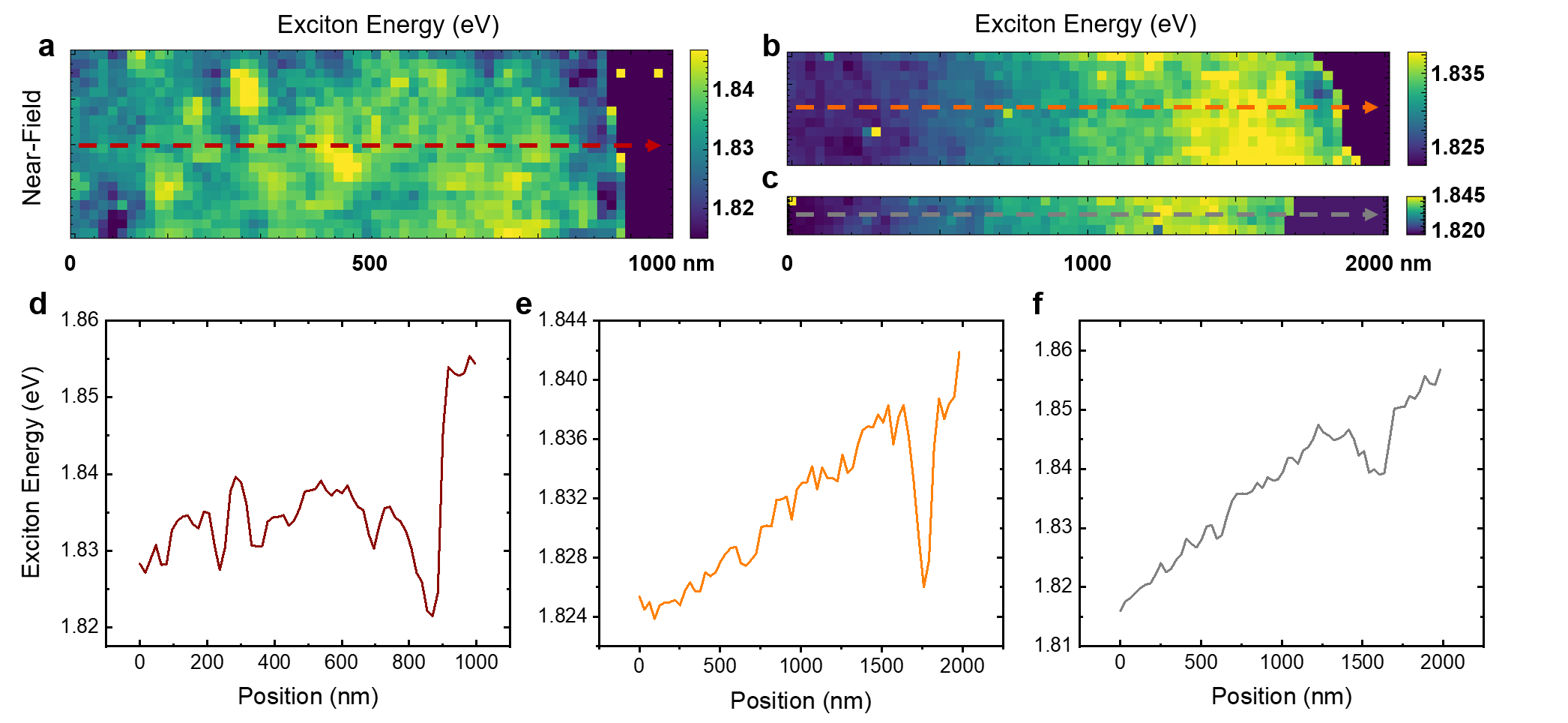} 
 \caption{\small {\bf a,c} Near-field exciton energy maps in other edge regions of MoS$_{2}$ monolayer sample A1 showing a similar PL redshift response presented at the edge region of Figure 4. {\bf d,f} Exciton energy profiles along the edges of ({\bf a,c}) highlighting the PL redshift feature. The profiles colors correspond to the dashed arrow colors of ({\bf a,c}).}
 \label{SI_Fig10}
 \end{figure}


\begin{figure}[!htb]
 \centering
 \includegraphics[scale=0.75]{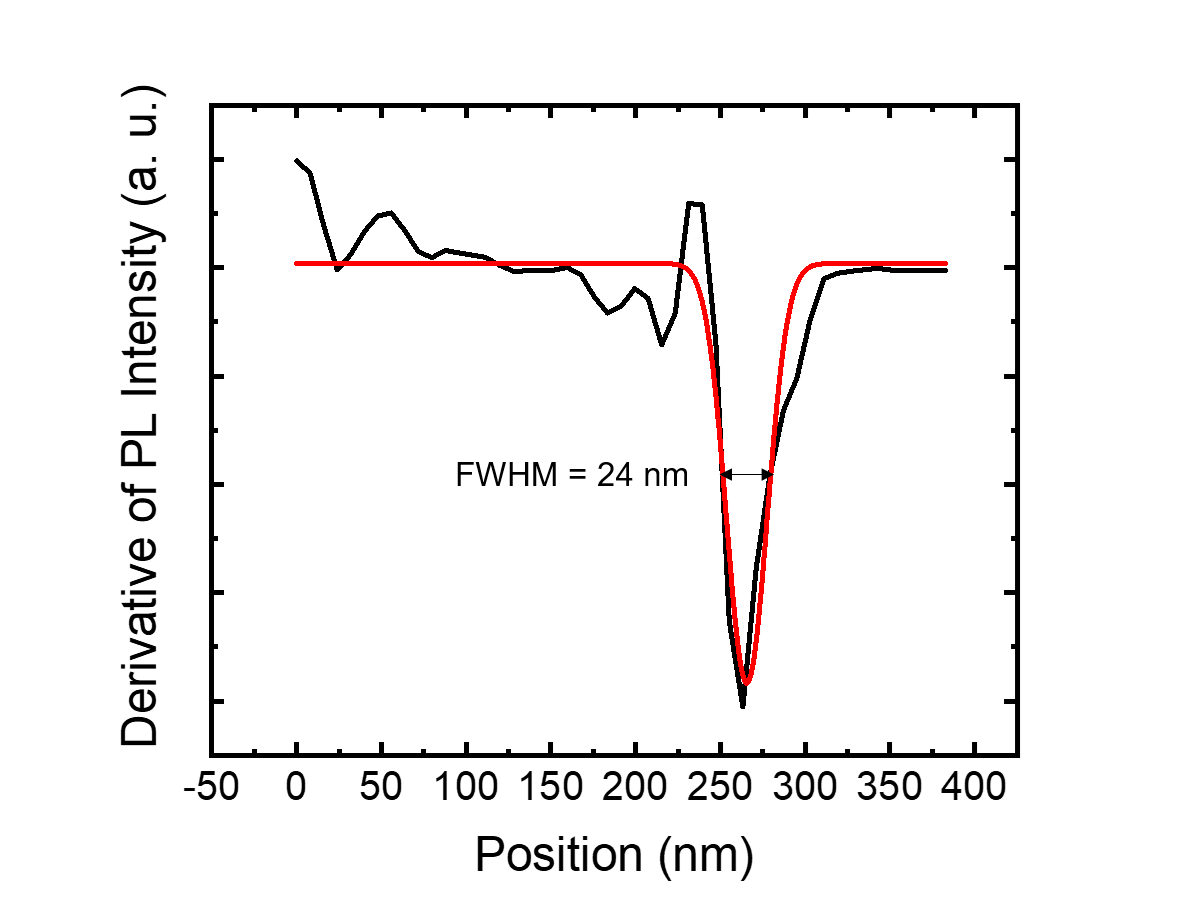} 
 \caption{\small To extract the spatial resolution of the near-field measurements we differentiated the PL intensity profile along the edge (in black) of Figure 4 and fitted it with a Gaussian function (in red). The spatial resolution of the measurement is the fitted full width at half maximum (FWHM) = 24 nm.}
 \label{SI_Fig11}
 \end{figure}


\begin{figure}[!htb]
 \centering
 \includegraphics[scale=0.7]{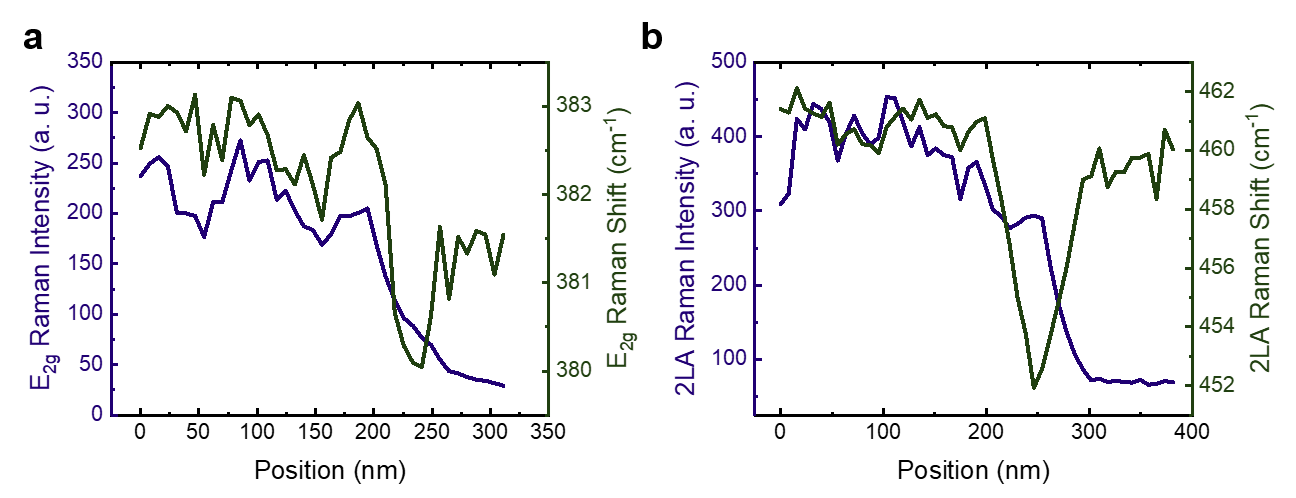} 
 \caption{\small {\bf a,b} E$_{2g}$ ({\bf a}) and 2LA ({\bf b}) intensity and frequency profiles along the edge of MoS$_{2}$ monolayer sample A1. These profiles were taken from the near-field Raman maps of Figure 5.}
 \label{SI_Fig12}
 \end{figure}


\begin{figure}[!htb]
 \centering
 \includegraphics[scale=0.5]{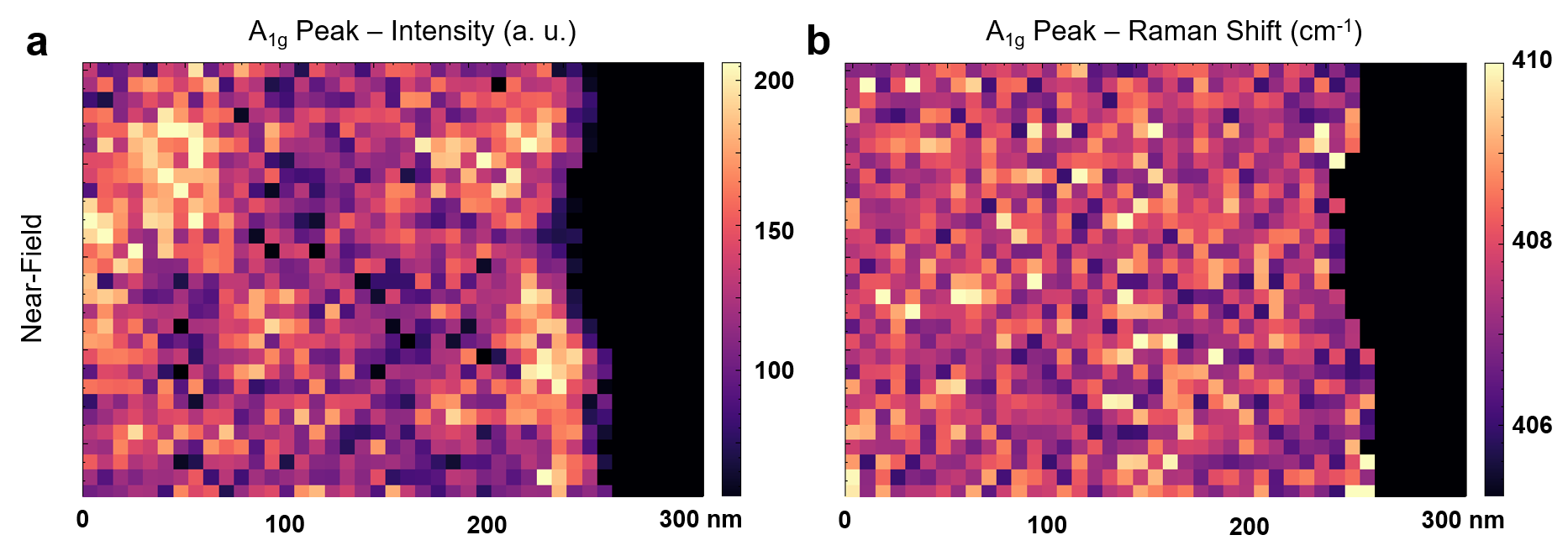} 
 \caption{\small {\bf a,b} Near-field A$_{1g}$ intensity ({\bf a}) and frequency ({\bf b}) maps at the edge of MoS$_{2}$ monolayer sample A1. These maps are from the same region of E$_{2g}$ and 2LA maps of Figure 5.}
 \label{SI_Fig13}
 \end{figure}

\end{document}